# Long-term resilience of online battle over vaccines and beyond

Lucia Illari[1]*, Nicholas J. Restrepo[2], Neil F. Johnson[1]

[1]Dynamic Online Networks Laboratory, George Washington University, Washington, D.C., 20052, U.S.A.

[2]ClustrX LLC, Washington, D.C., U.S.A.

**What has been the impact of the enormous amounts of time, effort and money spent promoting pro-vaccine science from pre-COVID-19 to now? We answer this using a unique mapping of online competition between pro- and anti-vaccination views among ~100M Facebook Page members, tracking 1,356 interconnected communities through platform interventions. Remarkably, the network's fundamental architecture shows no change: the isolation of established expertise and the symbiosis of anti and mainstream neutral communities persist. This means that even if the same time, effort and money continue to be spent, nothing will likely change. The reason for this resilience lies in "glocal" evolution: Communities blend multiple topics while bridging neighborhood-level to international scales, creating redundant pathways that transcend categorical targeting. The solution going forward is to focus on the system's network. We show how network engineering approaches can achieve opinion moderation without content removal, representing a paradigm shift from suppression towards structural interventions.**

## Introduction

Global confidence in vaccines has reached a critical inflection point despite unprecedented efforts to combat vaccine hesitancy[1–8]. Since the start of the COVID-19 pandemic, massive resources have been invested in content moderation, fact-checking, and behavioral science campaigns to overcome vaccine distrust. Facebook alone removed over 3,000 accounts and 20 million pieces of vaccine misinformation content[9–11], while initiatives like the Mercury Project committed $10 million to evidence-based counter-messaging campaigns[12,13]. The National Institutes of Health launched dozens of research grants aimed at understanding vaccine hesitancy and developing targeted interventions[14–17]. Yet vaccine skepticism remains at historic highs. A comprehensive study spanning 55 countries revealed that perceptions of vaccines' importance for children declined in 52 nations during the COVID-19 pandemic, with some countries experiencing drops up to 44 percentage points[18–20], , effects amplified by experimental evidence showing misinformation exposure reduces vaccination intent[21]. In the United States, approximately 20% of parents now express vaccine hesitancy[18,22]. Most alarmingly, vaccine-preventable diseases have resurged: the 2025 measles outbreak has claimed three unvaccinated lives and reached nearly 1,300 confirmed cases in the US[18,23,24] while Canada experienced over 3,800 cases[25]—nearly three times the US total despite having one-tenth the population. (For a comprehensive discussion of the global vaccine hesitancy crisis, the dismantling of research infrastructure, cross-pollination with other conspiracy narratives, and detailed analysis of intervention efforts, see Supplementary Material (SM) §2.)

The persistent failure of these comprehensive efforts reveals a fundamental misunderstanding about the nature of online misinformation. Current approaches treat misinformation as a content problem requiring content solutions: remove problematic sources, elevate credible voices, and ensure accurate information reaches audiences[9–17,26–48]. But online misinformation is not primarily a content problem—it is a network engineering problem. The issue isn't the presence of individual bad actors spreading false claims, but rather the structural properties of networks that enable certain types of information to flow efficiently to mainstream audiences while others remain confined to the periphery. Indeed, vaccine hesitancy correlates with network position within online communities rather than mere exposure to content[49]. When Facebook removes an anti-vaccination page, the assumption is that its influence disappears with it. But networks are complex systems with emergent properties, redundant pathways, and adaptive capacities that transcend individual components, and it *trust*, not content, that is the network's binding agent—rumors stick because the underlying relationships remain intact[50,51].

Here we show why content-focused interventions have failed and will continue to fail by conducting a longitudinal network analysis of Facebook's vaccine ecosystem from 2019 through 2025, tracking how 1,356 interconnected communities encompassing 70–100 million users evolved in response to platform interventions. We demonstrate the structural persistence of misinformation networks despite massive platform interventions: despite Facebook removing one-third of anti-vaccine pages and over half their connections, the network's functional architecture remains visually and structurally intact. We identify how the network evolved defensive adaptations that explain this resilience through topic and geographic "glocality" (i.e., pages increasingly blend vaccine discussions with climate change, elections, and other topics while connecting local communities to global conspiracy networks). But we also show how these same network properties can be leveraged for positive outcomes, including extreme opinion softening through small mixed-opinion deliberation groups that can achieve large-scale opinion moderation within weeks. Our analysis hence explains why current content-focused approaches have failed and instead points toward effective network engineering strategies for managing health misinformation at scale.

# Results

## System-level persistence despite volumetric losses

To understand why content-focused interventions have failed to meaningfully impact vaccine misinformation networks, we conducted a longitudinal analysis of Facebook's vaccine ecosystem from 2019 through 2025. Our dataset encompasses 1,356 interconnected communities representing 70–100 million users[52], tracked through multiple temporal snapshots that capture both the pre-pandemic baseline and the aftermath of extensive platform interventions. Each community represents a Facebook Page classified as pro-vaccine, anti-vaccine, or one of 12 neutral subcategories, with connections established when pages follow or like each other (see Methods/SM §1 for details).

The longitudinal comparison reveals a striking paradox: massive node and edge removal accompanied by fundamental structural preservation. Between November 2019 and June 2025, Facebook's vaccine ecosystem experienced substantial losses across all categories. Anti-vaccination pages suffered heaviest attrition (168 of 501 pages, 33.5% removed), followed by neutral pages (177 of 644, 27.5%) and pro-vaccination pages (37 of 211, 17.5%). Connection losses proved even more dramatic: anti-vaccination pages lost 52.3% of edges, neutral pages 48.5%, and pro-vaccination pages 33.8%. This represents systematic degradation of network connectivity, with anti-vaccination and neutral communities bearing the brunt of dissolution. Yet despite these substantial volumetric losses—approximately 28% of all pages and 47% of all connections—the network's fundamental architecture remains visually and structurally intact (**Figure 1**A–B). The ForceAtlas2 layout visualizes structural relationships as spatial arrangements

(see Methods/SM §1 for details)[53–56]. Our objective is not to overinterpret specific node positions, but to demonstrate how clusters maintain their essential organizational logic despite massive node and edge removal. The spatial positioning reveals that broader topological patterns persist unchanged across the intervention period.

**Figure 1** shows that despite enormous investments to dampen anti-vaccination influence, there is no visual change in the system-level structure. The network remains robust and resilient at meso- and macroscale levels. This structural persistence was maintained consistently throughout the intervention period—the network could in principle have undergone degradation and recovery cycles as observed in other competitive social networks under external pressure[57] to show the form in **Figure 1**B, but it did not, instead maintaining its essential form continuously. The layouts reveal that while individual nodes changed, the organizational logic persists: anti-vaccination pages continue forming dense red hubs entangled with neutral Facebook pages, while pro-vaccination clusters remain peripheral. The tri-polar geometry with neutral bridges demonstrates remarkable structural stability across five years. The corollary is profound: future investments along the same lines will almost certainly have no significant effect on the network's fundamental structure or capacity to facilitate anti-vaccination influence. Instead, the system-level structure across microscale, mesoscale and macroscale must be the focus of future efforts, combined with understanding topic blends and geographic distributions within individual pages.

The persistence of network activity further confirms that structural preservation translates into functional continuity. Analysis of first-page posting activity from January through June 2025 (**Figure 1**C) demonstrates that surviving Pages remain actively engaged rather than serving as dormant archives. This measure captures the recency of the most recent ~10 posts visible on each Page's first screen, providing an indirect but revealing snapshot of ongoing network vitality. Pages with high posting frequencies will concentrate their activity in recent months (appearing primarily in the June column), while less active pages distribute their limited posts across the timeline as they post sporadically throughout the period. The resulting distribution reveals two key insights: first, that the network continues to function as a living information ecosystem rather than a collection of abandoned or zombie pages, and second, that all three stance categories—anti-vaccination, neutral, and pro-vaccination— maintain active posting communities. The breakdown by stance category shows that despite the volumetric losses in nodes and edges, each segment of the network retains pages that are regularly updating their content, indicating that the network's reduced size has not compromised its basic capacity for ongoing information generation and circulation across the ideological spectrum.

This structural resilience reveals a fundamental scale mismatch in current mitigation strategies. Traditional removal approaches assume networks function as simple aggregation systems where influence scales proportionally with active nodes and edges. While content moderation and fact-checking help individual users, our findings suggest misinformation networks are complex adaptive systems with emergent properties, redundant pathways, and robust architectural principles that persist across multiple disruption scales. **Figure 1** demonstrates that individual-level interventions cannot address network-level dynamics enabling vaccine hesitancy to persist at population scale. Deleting false content cannot, by itself, neutralize hesitancy because the relational scaffolding of trust endures beyond any single post[50,51]. Tackling misinformation at this scale requires fundamentally different strategies that work with rather than against underlying network architecture.

## Glocal evolution as adaptive defense

The resilience observed in **Figure 1** emerges from a fundamental transformation in how vaccine-related discourse operates within Facebook's ecosystem. **Figure 2** reveals that the debate has expanded far beyond traditional vaccine topics to create "topic glocality"—seamless integration of diverse subjects that renders targeted content moderation ineffective. **Figure 2**A shows the same 2025 network from Figure 1B, with nodes colored by 14 distinct topic and stance categories: anti-vaccine (red), pro-vaccine (blue), and 12 neutral subcommunities spanning parenting, alternative health, GMO discussions, social movements, and others. The black circle marks where most Facebook-targeted anti-vaccine pages once operated, yet surviving pages form a tightly woven fabric spanning multiple topical boundaries, creating an information ecosystem where no subject domain can be cleanly isolated or targeted.

The zoom-in panels (**Figure 2**B–F) reveal the mechanism underlying this resilience: individual pages routinely discuss several ostensibly unrelated topics, as illustrated by the mix of colored wedges within single nodes. A parenting page might simultaneously post about climate change and elections, while an alternative health community discusses abortion rights alongside vaccine concerns. This topic agnosticism creates what effectively functions as a multi-lane information highway where content and audiences traverse issue boundaries with remarkable ease. The gray inter-topic connections visible in the zoom-ins demonstrate further cross-pollination between communities, showing how discussions initiated in one topical domain can rapidly spread across the entire network regardless of the originating community's stated focus.

This topical diversification represents a critical adaptive defense mechanism emerging from platform interventions. Rather than maintaining discrete silos that could be individually targeted, the network evolved to integrate diverse topics within and across communities. Traditional content moderation approaches designed for specific topics now encounter a moving target spanning abortion rights, climate change, elections, parenting advice, and health alternatives simultaneously. When platforms remove topic-focused content or communities, the network's topic-agnostic structure preserves underlying audience relationships and influence pathways by shifting conversations to adjacent topics while maintaining the same skeptical worldview across domains. These findings echo actor-level work mapping the "medfluencer" ecosystem on X, where the same follower graph carries a shifting repertoire of topics[58]; both results underscore that once trust ties form, content can mutate without eroding reach.

The topic glocality in **Figure 2** operates alongside an equally important geographic dimension strengthening network resilience. **Figure 3** demonstrates how the Facebook vaccine ecosystem transcends spatial boundaries through "geographic glocality"—seamless integration of local communities with global networks that renders geographically-targeted interventions ineffective. **Figure 3**A reveals the global footprint with red dots representing pages' declared locations and white lines showing inter-page connections, geocoded at centroid level. Dense connections crisscrossing continents illustrate how local discussions routinely bridge to global audiences, creating pathways from neighborhood communities to worldwide hubs without geographic constraints. **Figure 3**B's East Coast inset shows that physical proximity does not equate to ideological similarity: anti-vaccine, neutral, and pro-vaccine pages coexist in the same regions, all densely interconnected.

**Figure 3**C's toponym co-occurrence map reveals the underlying mechanism: place names scraped from page titles create a lattice binding different spatial scales. Eight color-coded label families span neighborhood to continent scale, with connections linking radically different geographic scopes: "Nashville TN" connects to "Global," "Brisbane" bridges to "Europe," "Canada" links to "Australia." This cross-scale wiring explains why deleting region-specific pages often fails to stem global misinformation flow: the network's geographic structure ensures that local narratives can immediately amplify through global channels, while worldwide movements can rapidly localize through community-specific pages. The result is an information ecosystem where attempts to contain misinformation within specific geographic boundaries encounter a network architecture explicitly designed to transcend such limitations, creating redundant pathways that span from hyperlocal neighborhood pages to continental and global hubs.

Together, **Figure 2** and **Figure 3** reveal how network resilience emerges from dual glocality operating simultaneously across topical and geographic dimensions. Topic agnosticism (**Figure 2**) works with geographic boundary-crossing (**Figure 3**) to create a robust information ecosystem with redundant pathways at multiple scales. When content moderation targets vaccine misinformation in specific domains, topic glocality ensures audience relationships persist through adjacent subjects like abortion rights or climate policy. When platforms attempt geographic containment by removing region-specific pages, geographic glocality provides alternative routes through cross-scale connections spanning neighborhood to continental networks. Traditional intervention strategies, whether topic-focused or geographically-targeted, encounter a moving target that instantly reconstitutes through alternative topical framings and geographic pathways. This dual glocality transforms the network from discrete, targetable communities into an integrated, adaptive system maintaining influence despite sustained platform interventions.

## Network engineering solutions

The structural insights revealed by our network analysis point toward a fundamentally different approach to managing vaccine misinformation: rather than fighting against the network's architecture through content removal, we can harness its existing properties to promote constructive dialogue and opinion moderation. **Figure 4** demonstrates the potential of network engineering—purposeful interventions that modify how information flows through existing network structures without removing nodes or suppressing content. The core mechanism involves creating small, mixed-opinion deliberation groups drawn from the network's existing neighborhood structure rather than randomly assembling strangers (see SM §9 for details). Pilot experiments involving over 100 participants[59] and large-scale AI-mediated deliberation studies with over 5,000 UK participants[60] demonstrate that people moderate their views when engaging with others sharing common ground or mutual connections. AI-mediated deliberation proved particularly effective, with participants' stances converging toward common positions and AI-generated group statements receiving higher endorsement than human-mediated alternatives[60]. Previous studies of AI chatbots for vaccine confidence have demonstrated some potential when deployed as standalone, non-networked tools[61].

Our agent-based simulations extend these empirical findings[59] to the current Facebook vaccine network (**Figure 1**B), modeling how repeated application of such deliberation methods could transform the ecosystem at scale. **Figure 4**A shows snapshots of the network transformation: starting from the original configuration with distinct red (anti-vaccine), blue (pro-vaccine), and green (neutral) communities, the bottom row of **Figure 4**A shows outcomes after repeated micro-deliberations. In the average-case scenario (**Figure 4**A bottom row-left), almost 90 anti-vaccine accounts shift to neutral positions, significantly moderating the network's overall stance distribution. The better-case scenario (**Figure 4**A bottom row-right) shows even more dramatic results when focusing intervention efforts specifically on anti-vaccine voices, yielding a substantially more neutral core while maintaining the network's essential connectivity.

The temporal dynamics of this transformation are equally striking, as shown in **Figure 4**B's conversion trajectories over simulation time-steps. The solid lines track the average number of remaining anti-vaccine (red) and pro-vaccine (blue) accounts across 1,000 independent simulation runs, with shaded bands indicating statistical confidence. The dashed vertical lines mark critical milestones: the average moment when 50% of each group has converted to neutral, and the grey band showing the mean time to achieve full network neutralization. These results suggest that systematic opinion moderation can occur within a matter of weeks—a timeframe supported by the empirical studies showing significant attitude shifts after single deliberation sessions[59].

The elegance of this network engineering approach lies in its alignment with, rather than opposition to, the system's natural dynamics. Instead of fragmenting the network by removing nodes and edges—a strategy our analysis showed to be ineffective—the deliberation method preserves existing relationships while gradually shifting opinion distributions flowing through them. This approach operates without censorship or content removal, addressing primary sources of backlash against platform interventions by providing additional opportunities for respectful cross-perspective dialogue within existing social networks. This creates a "democratic moderation"[60] where opinion change emerges from voluntary participation rather than top-down content control. This represents a paradigm shift from suppression-based to transformation-based interventions, with empirical evidence suggesting network engineering could achieve the opinion moderation that traditional content-focused approaches have failed to deliver while avoiding polarization and resistance accompanying censorship-based interventions.

## Discussion

Our analysis of Facebook's vaccine ecosystem reveals a fundamental mismatch between current intervention strategies and the structural realities of online misinformation networks. Despite losing approximately one-third of anti-vaccine pages and over half their connections, the network's tri-polar architecture persisted unchanged across five years of intensive platform interventions. This resilience stems from the network's adaptive capacity to maintain functional pathways despite node removal through "glocal" evolution: Pages increasingly blend vaccine discussions with climate change, elections, and local community concerns while bridging hyperlocal neighborhoods to global conspiracy networks. This dual boundary-crossing renders traditional categorization-based interventions ineffectual, as content appearing to be about parenting or local politics can serve as conduits for vaccine misinformation.

The scale mismatch between intervention strategies and network dynamics represents the most critical oversight in current approaches. Individual-level interventions (for e.g., removing pages, fact-checking posts, targeting users with corrective messaging) cannot address the network-level properties that enable systematic misinformation flow. This is exemplified by pro-vaccine communities' persistence at the network periphery: despite retaining roughly 80% of their presence, they failed to challenge anti-vaccine hubs' central role, illustrating how structural positioning matters more than raw capacity. The solution lies not in creating more pro-vaccine content or removing more anti-vaccine pages, but in engineering information flows that leverage existing topology for cross-perspective dialogue. Our findings demonstrate that the same structural features conferring resilience against platform removals can be harnessed for constructive outcomes through network engineering, aligning with frameworks emphasizing that community-level resilience requires structural approaches[62]. Rather than fragmenting networks by removing nodes, deliberation methods preserve existing relationships while gradually shifting opinion distributions flowing through them. Empirical evidence from autonomous online deliberation and AI-mediated group discussions shows people readily moderate views when engaging in structured dialogue within existing social networks, achieving the opinion softening that years of content removal have failed to deliver.

These findings extend beyond vaccine misinformation to other polarized ecosystems where dual glocality enables adaptive resilience, suggesting a paradigm shift from content suppression to network topology interventions. AI-mediated deliberation systems could be integrated into platform infrastructures to create ongoing cross-perspective dialogue opportunities, while network analysis tools could identify optimal intervention points. However, such approaches require moving beyond platform governance as content policing toward understanding networks as complex adaptive systems. The ultimate test—population-scale validation of micro-deliberation effects—lies beyond this work's scope but would represent a fundamental advance in fostering democratic dialogue in our polarized digital age.

## Methods

### Data Collection and Page Classification

Our methodologies for data collection, classification and network construction in Fig. 1 are an extension of earlier works[52,63,64]. We provide pictures of sample Facebook Pages in the Supplementary Material. Specifically—and exactly as in previous studies—each node in the network in **Figure 1** represents a community of members of a given Facebook Page that has become connected into the online competition between pro- and anti-vaccination views. Indeed, **Figure 1**A is an essentially identical diagram to that earlier work[52] apart from minor cosmetic differences due to re-running the ForceAtlas2 layout. We stress that all our data were collected from public Facebook Pages representing organizations, causes, communities, or public figures. These Pages are distinct from personal Facebook accounts, which represent private individuals. We did not collect data from personal accounts or private groups, and none of our analysis required access to individual-level information.

Our analysis focuses on the ecology of Facebook Pages because recent studies confirm that people (for e.g., parents) share their concerns and views with others largely through such built-in communities, including extreme views[65–67]. Though for simplicity we refer to each Facebook Page as a community, we stress that it is unrelated to any ad hoc community structure inferred from network algorithms. Each Page aggregates people around some common interest, is publicly visible, and its analysis does not require access to personal information.

Each Page can also provide a link to another Facebook Page whose content is of interest to it. Representing each Facebook Page as a node, nodes are connected by directional links when Page *A* follows or likes Page *B*, meaning Page *A* explicitly listed Page *B* as one of the Pages to which it linked, not necessarily due to agreement, but because Page *B*'s content was of interest to Page *A*'s community. The important consequence of such a link from Page *A* to Page *B* is that *B*'s content (which can include any extreme views) then tends to appear in community *A*'s content feeds, creating an information conduit through which *B* can influence *A*. Although other definitions of nodes and links are possible, our chosen approach negates the need for individual-level information and uniquely defines each community (node) since each Page possesses a unique identification number.

To obtain the relevant Facebook Pages (nodes) in **Figure 1**, we proceeded as follows. We constructed the Facebook vaccine ecosystem network through iterative snowball sampling starting with manually identified seed Pages discussing vaccines, vaccination policies, or the pro-vs-anti vaccination debate, obtained by searching Facebook's platform in 2018-2019. Using Facebook's GraphAPI, we captured outbound "follow" or "like" links between Pages (established at the administrator level, not individual user posts) and iteratively expanded the network by adding newly discovered connected Pages. We retained Pages that either discussed vaccines directly or self-identified as causes, communities, or NGOs connected to vaccine-related discourse, manually pruning connections at each iteration to extract meaningful relationships rather than creating a nearly fully connected network. This process yielded 1,356 interconnected Facebook Pages representing approximately 86.7 million individuals across multiple countries and languages.

Our trained researchers classified each Page as pro-vaccine, anti-vaccine, or neutral based on manual review of recent posts, page descriptions, and self-identification categories. Pro- and anti-vaccine classifications required either 2+ of the most recent 25 posts addressing the vaccine debate or explicit stance identification in page titles/descriptions. Neutral pages required vaccine-related content without explicit stance-taking or self-identification as NGOs, causes, or communities connected to vaccine discourse, with further sub-categorization into 12 topic areas (parenting, alternative health, GMO discussions, social movements, etc.). Subject matter experts classified each page independently with disagreements (approximately 15% of cases) resolved through team discussion until consensus. We included only Pages in languages understood by our multilingual research team, such as English, French, Spanish, Italian, Dutch, and Russian. Complete methodological details, classification criteria, and validation procedures are provided in Supplementary Material §1.

## Network Mapping and Content Classification

This process yielded a network of 1,356 interlinked Facebook Pages comprising approximately 86.7 million individuals across countries and languages: 211 pro-vaccine Pages (blue nodes) with 13.0 million individuals, 501 anti-vaccine Pages (red nodes) with 7.5 million individuals, and 644 neutral Pages (various colored nodes) with 66.2 million individuals. We can estimate a size for each community by its number of likes (fans) because a typical user only likes 1 Facebook page on average[68]: this size typically ranges from a few hundred to a few million users, but we stress that our analysis and conclusions do not rely on us determining community sizes. The public information that we gather about each page's managers[69] suggests that users come from a wide variety of countries (see SM §3 for details). The most frequent manager locations are the United States, Australia, Canada, the United Kingdom, Italy, and France (SM §3).

The resulting network is scalable to the population level and visually manageable when analyzed using the open-source software Gephi with its ForceAtlas2 algorithm[53–56]. To present the resulting networks in **Figure 1**A–B, **Figure 2** A–F, and **Figure 4**A, we employ the ForceAtlas2 network layout, a layout mechanism where nodes repel each other while links act like springs, bringing interlinked nodes closer together[55]. Hence, closer nodes are more likely to share and discuss similar content, while distant nodes are not. The layout is agnostic to color differentiation: the segregation of colors, or node categories, is an entirely spontaneous effect, independent of the algorithm.

To analyze the evolution of discourse beyond vaccines, we categorized the content within each of the 1,356 communities by topic prevalence. While numerous topics appear in community discussions, five emerged as dominant: COVID-19, mpox, abortion, elections, and climate change. We developed keyword filters to identify discourse around these five non-vaccine topics, where "non-vaccine" refers to topics not centered on vaccines in a broad sense. Although our COVID-19 and mpox filters necessarily included some vaccination-related terms (such as "monkeypox vax'n"), our intent was to capture disease-specific discourse rather than general vaccine discussion. This distinction was important because filtering for general vaccine content would have made it difficult to reliably distinguish COVID-19-related posts from mpox-related posts through automated methods without extensive additional human classification. The filters employed regular expressions and keyword searches applied across post content, page descriptions, image tags, and link text in multiple languages.

Our analysis revealed important geographic and topical dimensions within the network structure. Of the 1,356 communities, 342 identify as geographically local (encompassing approximately 3.1 million individuals), while the remaining 1,014 communities operate at a global scale (encompassing approximately 83.7 million individuals). We apply the terms "global" and "local" across both geographic and topical dimensions to examine the interconnection between what we term geographic and topic "glocality." This dual application illuminates how geographic and topic glocality can interrelate, allowing communities to occupy different glocal positions along both dimensions simultaneously. For comprehensive details on topic filtering methodology, statistical validation of topic-geographic relationships through chi-square tests, temporal analysis of cross-topic discourse patterns, and quantitative assessment of glocality correlations with network centrality measures, see Supplementary Materials §6–8.

Facebook amplified its interventions around November 2020, targeting vaccines and COVID-19 content[70]. We obtained network snapshots before, during, and after this intervention period, with our baseline from November 2019 (**Figure 1**A) and updated mapping through 2025 (**Figure 1**B). Since 2019, this nearly 100 million user ecology broadened from vaccines to COVID-19 policies and subsequently to climate change, elections, abortion rights, and local community issues, demonstrating the "glocal" evolution described in our results (see SM §5 for pre-intervention content examples).

## Data Availability

The datasets used in this study contain sensitive information from social media platforms. To comply with data protection standards and avoid potential misuse, the raw data cannot be shared publicly, however the preprocessed derivative datasets which can be used to reproduce the results in the study are available in our Data Access repository, https://github.com/gwdonlab/data-access.

## Code Availability

The code used to generate the networks in **Figure 1**A–B, **Figure 2**A–F, **Figure 3**C, and **Figure 4**A is Gephi, which is free open-source software. **Figure 1**C was obtained using the open-source graphing library Plotly. **Figure 3**A and **Figure 4**B were obtained using Mathematica.

## Acknowledgements

This research received no specific grant from any funding agency in the public, commercial or not-for-profit sectors.

## Author contributions statement

L.I. analyzed the results and generated the figures. N.F.J. supervised the project. L.I. and N.F.J. wrote the paper. All authors were involved in reviewing the final manuscript, and in the conceptualization, methodology, and validation.

**Competing Interests**

The authors have no competing interests, either financial and/or non-financial, in relation to the work described in this paper.

**Corresponding Author**

All correspondence and material requests should be addressed to L.I. loi2102@gwu.edu

**Figures**

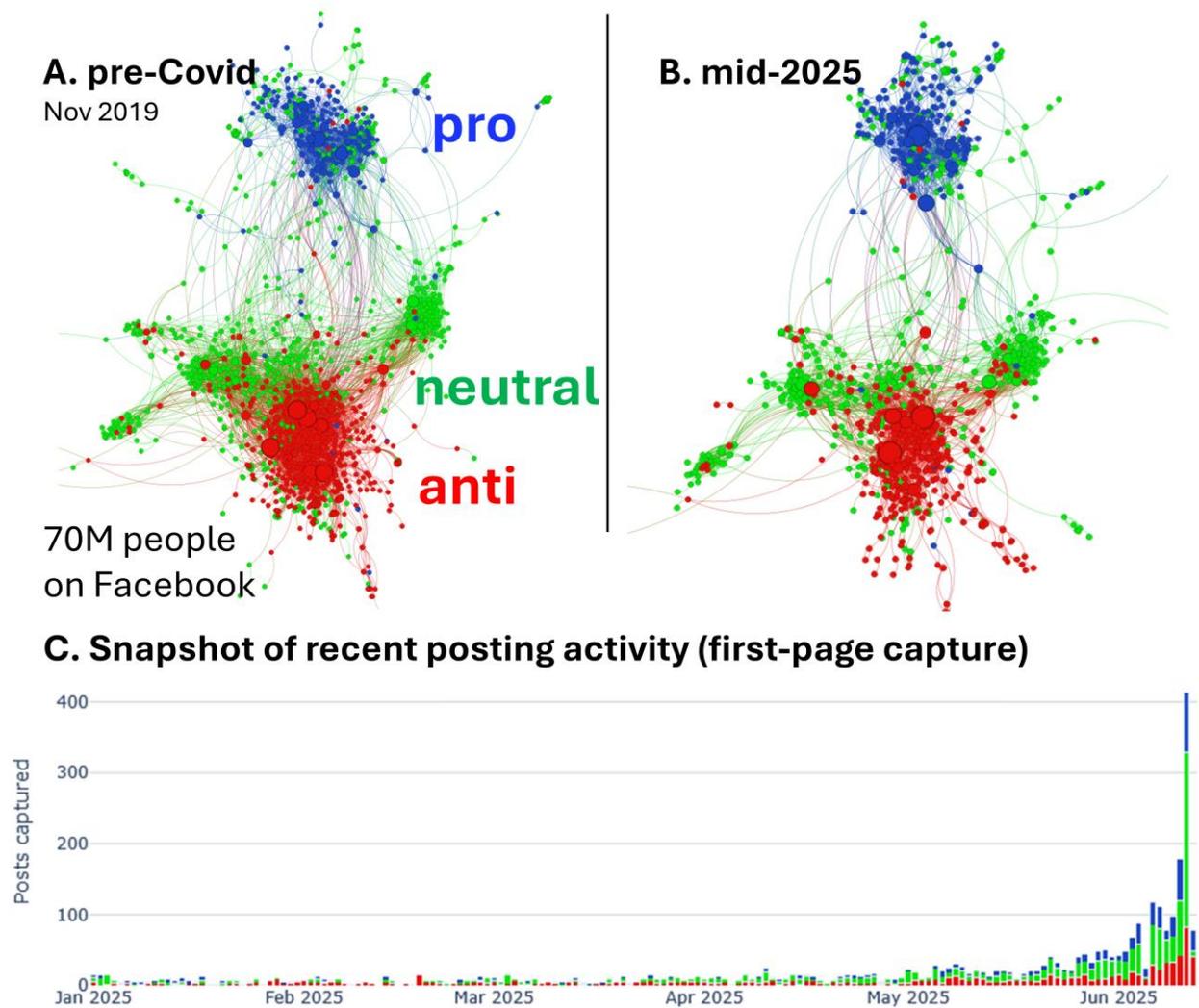

*Figure 1. The Facebook vaccine ecosystem is visually unchanged after six years of interventions.* (A) Pre-COVID snapshot, Nov 2019. (B) Live pages as of 11 Jun 2025. Node size = betweenness centrality; colors denote stance. Despite the removal of 33% of anti-vax pages (red) and 28% of all pages, the anti-neutral-pro geometry is intact. (C) Recent posting activity of all live pages (first-page capture, Jan–Jun 2025) confirms that the network remains active, not archival.

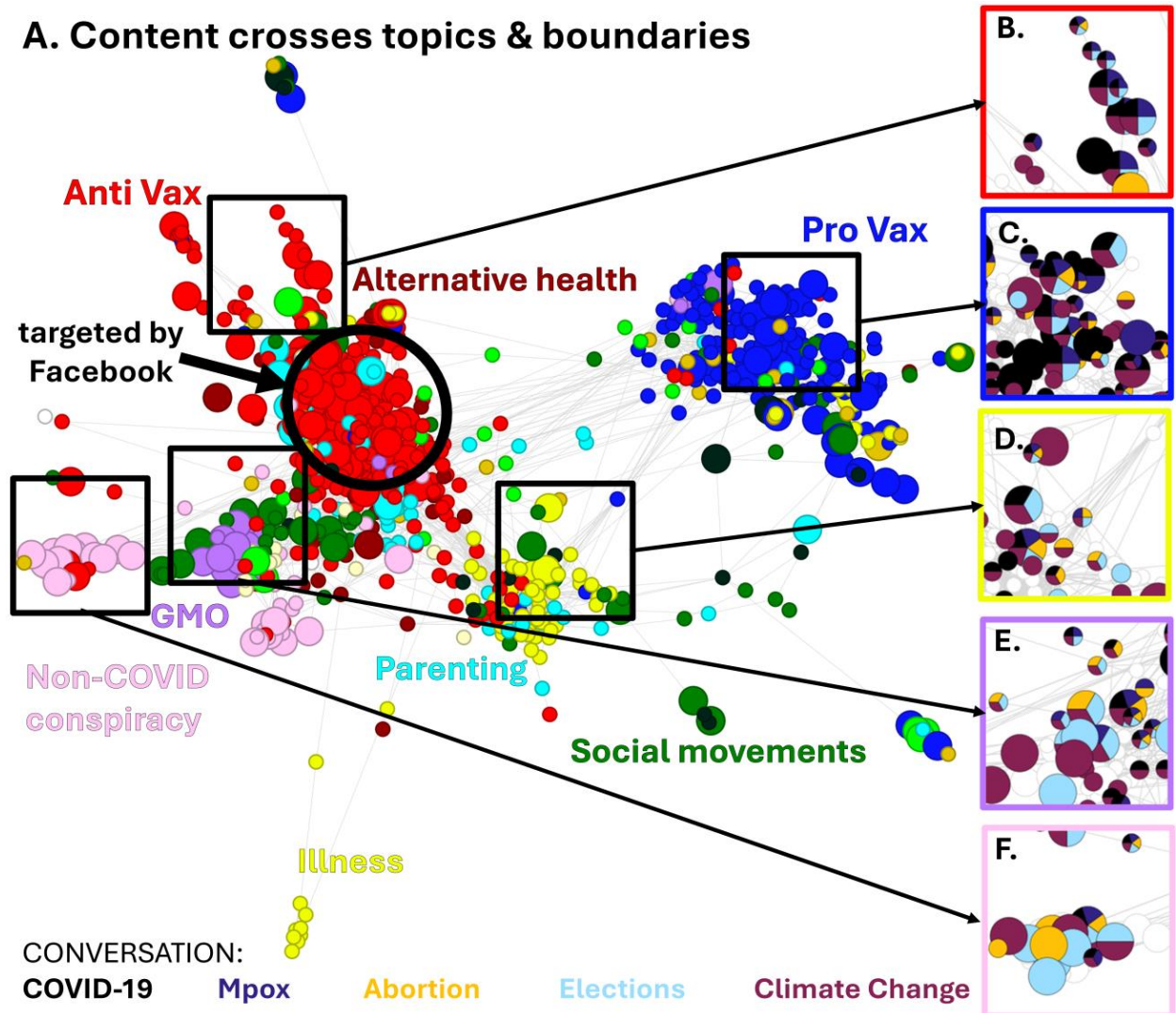

***Figure 2. Topic-level "glocality" keeps vaccine debate resilient.*** *(A) 2025 Facebook vaccine network from Fig. 1B, but nodes are now colored by 14 topic/stance categories: Anti-vaccine (red), Pro-vaccine (blue) and 12 Neutral sub-communities (greens, yellows, purples, etc.). Node size = betweenness centrality. A black circle marks the zone in which most of the Anti-vaccine pages removed by Facebook once sat. Despite those removals, the surviving pages form a tightly woven fabric that spans multiple topics (e.g. "Parenting", "Alternative health", "GMO", "Social movements") so that no subject silo can be cleanly isolated. (B–F) Zoom-ins on five representative regions (boxes in A). Each panel shows an eclectic mix of colored wedges inside individual nodes, illustrating how single pages routinely discuss several ostensibly unrelated topics (for e.g. a Parenting page posting about climate change and elections). Grey inter-topic edges reveal further cross-pollination. Together the panels demonstrate a topic-agnostic information highway: content and audiences traverse issue boundaries with ease, thwarting moderation strategies that target only one subject domain.*

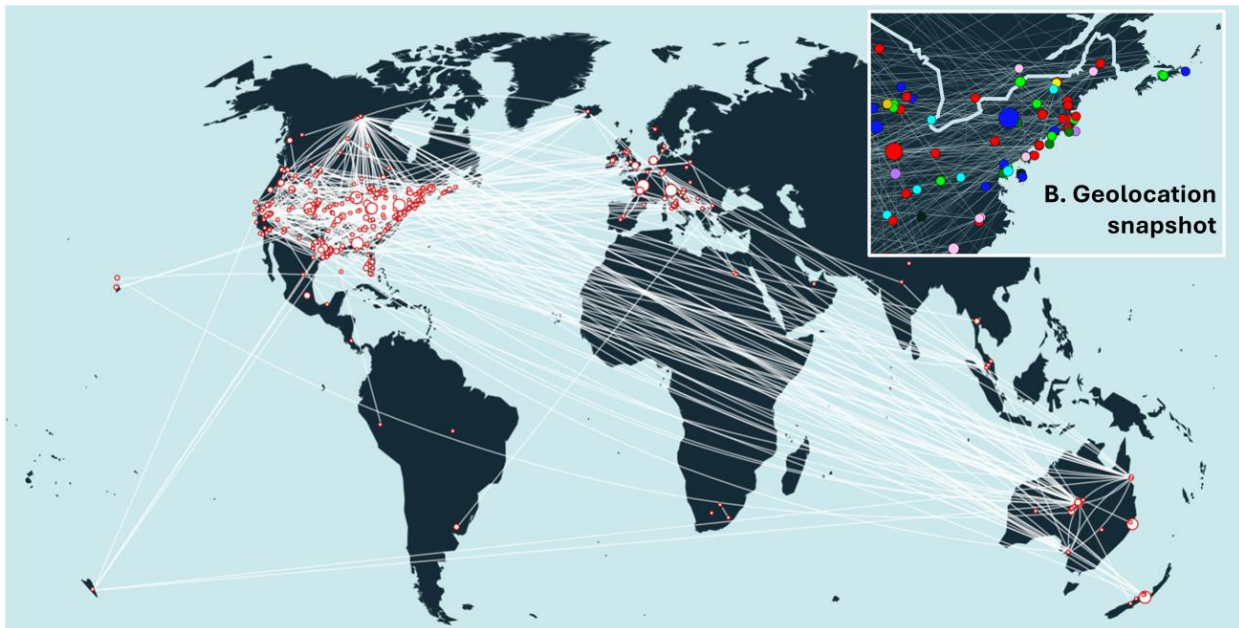
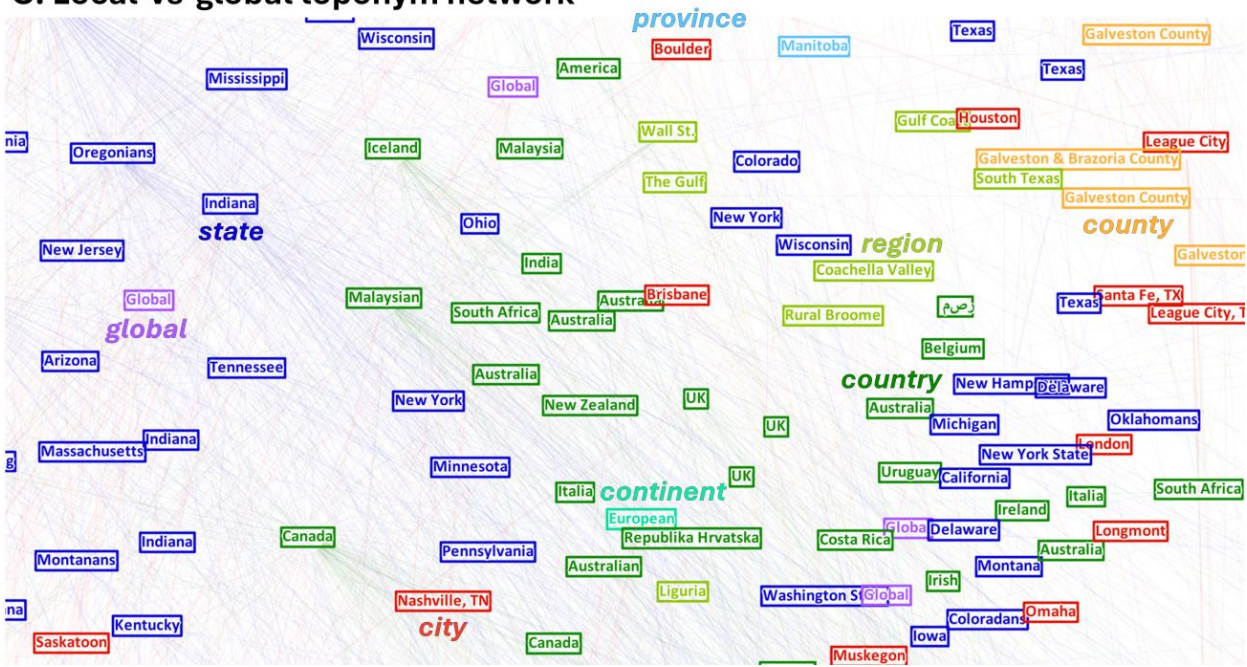

*Figure 3. Geographic "glocality" enables cross-scale spread of misinformation.* (A) Geographic footprint of the same Facebook vaccine ecology. Red dots = pages with declared locations; white lines = inter-page connections geocoded at centroid level. Connections crisscross continents, indicating that local discussions routinely bridge to global audiences. (B) East-Coast inset (box in A) colored by stance, underscoring that physical proximity does not equate to ideological similarity; anti, neutral, and pro-vaccine pages lie side-by-side, all densely interconnected. (C) "Local-vs-Global" toponym co-occurrence map. Labels in eight color families mark place names scraped from page titles or 'About' sections (neighborhood to continent scale). Links connect pages whose titles mention different spatial scales, for e.g. "Nashville TN" ↔ "Global", "Brisbane" ↔ "Europe", revealing a lattice that binds

*neighborhood-level communities to worldwide hubs. This cross-scale wiring explains why deleting region-specific pages often fails to stem global misinformation flow.*

## A. Network softening snapshots

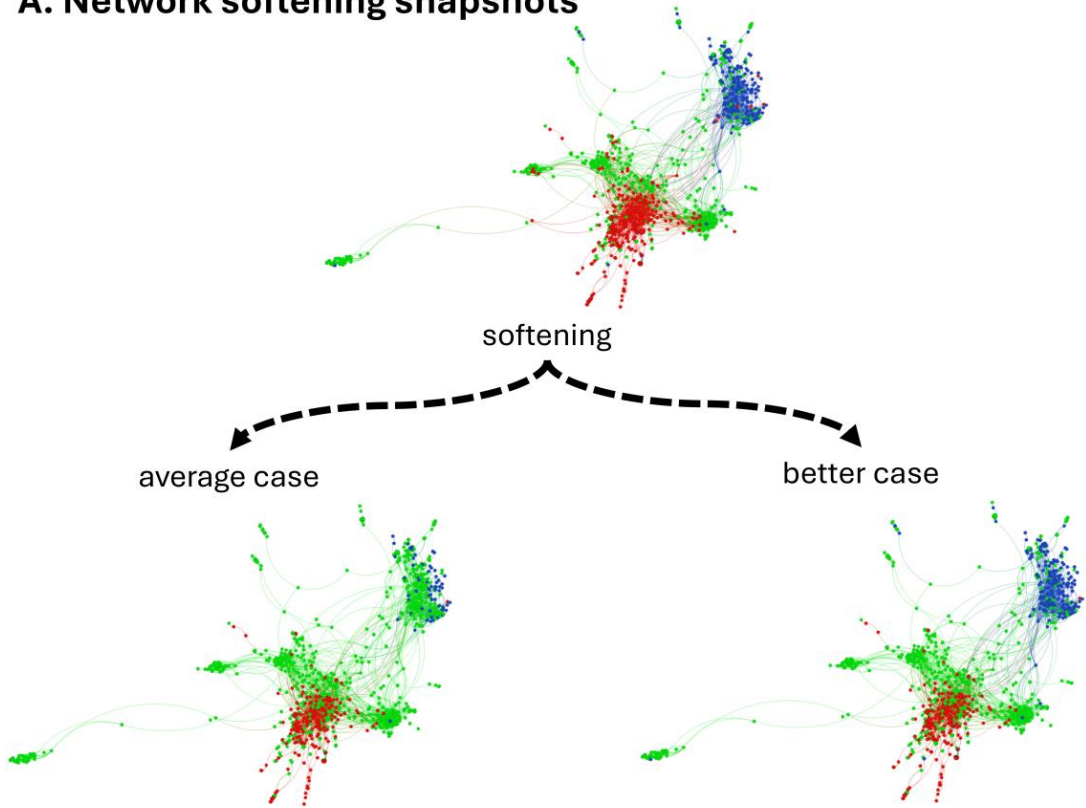

## B. Conversion trajectories over time

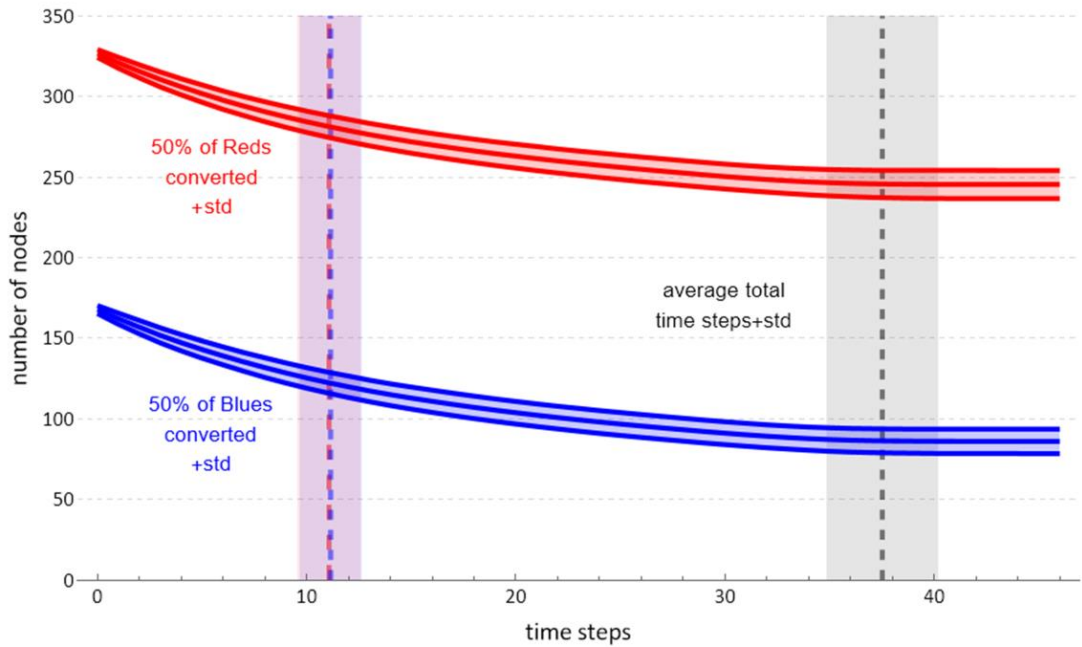

*Figure 4. Simulated "softening" shows how small mixed-opinion circles can disarm extremes.* (A) Snapshots of agent-based softening simulations run on the June 2025 live network (node colors as in Fig. 1). Top: starting configuration. Bottom-left: average-case outcome after repeated one-hour micro-deliberations in randomly drawn neighborhood circles; roughly one-third of Anti-vaccine (red)

*accounts shift to neutral (green). Bottom-right: better-case scenario in which only Anti-vaccine voices are eligible to change, yielding a cleaner neutral core. (B) Mean conversion trajectories (solid lines) ±1 s.d. (shaded bands) over simulation time-steps. Red and blue dashed lines (overlapping) mark the average moment when 50 % of Anti- and Pro-vaccine nodes, respectively, have become neutral; the grey band shows the mean time to full neutralization. Although Reds initially outnumber Blues, both curves decay with the same characteristic shape, because conversion probability depends solely on landing in a mixed circle. The experiment demonstrates that network-aware, bottom-up dialogue mechanisms can achieve large-scale moderation in weeks—without removing pages or suppressing content. (Model specification and parameter sweeps in SM §9.)*